\documentclass[aps,prl,twocolumn]{revtex4}
\begin{document}
\title{String theory T-duality and the zero point length of spacetime}
\author{A.~Smailagic$^\ast$, E.~Spallucci$^\dagger$, T.~Padmanabhan$^\ddagger$}

\affiliation{Sezione INFN di Trieste,\\
         Strada Costiera 11, 34014 Trieste,\\
         Italy}
\email{e-mail: anais@ictp.trieste.it}

\affiliation{Department of Theoretical Physics,\\
         University of Trieste, Strada Costiera 11, 34014 Trieste,\\
         Italy, and\\ Sezione INFN di Trieste,\\
         Strada Costiera 11, 34014 Trieste,\\
         Italy}
\email{e-mail: spallucci@trieste.infn.it}
\homepage{http://www-dft.ts.infn.it/~euro/}

\affiliation{IUCAA, 
Post Bag 4, Ganeshkhind, Pune - 411 007}
\email{e-mail: nabhan@iucaa.ernet.in}
\homepage{http://www.iucaa.ernet.in/~paddy}

\date{\today}
\begin{abstract}
It has been often conjectured that the correct theory of quantum gravity will 
act as a UV regulator in the low energy limit of quantum field theory. 
Earlier work has shown that if the path integral defining the quantum field 
theory propagator is modified, so that the amplitude is invariant under the
duality transformation $l\to 1/l$ where $l$ is the length of the path, 
then the propagator is UV-finite and exhibits a ``zero-point length'' of the 
spacetime. Since string theory uses extended structures and has a T-duality, 
these results should also emerge directly from string theory. We show, by
explicit path integral computation, that this is indeed the case. 
The lowest order string theory correction to the propagator 
is the same as that obtained by the hypothesis of path integral duality.
\end{abstract}

\maketitle

\noindent
Conventional quantum field theory exhibits UV divergences and it was always 
thought that this behavior will be
``cured'' in the correct theory of quantum gravity  \cite{garay}. This is to 
be expected since one cannot operationally define
scales below Planck scale, in any model that incorporates the principles of 
quantum theory and gravity \cite{padma1},
making Planck length act as the ``zero-point length'' of the spacetime. 
The propagator in a field theory should therefore 
be ultraviolet finite when quantum gravitational corrections are incorporated.
 
It was shown in \cite{padma2} that
if the path integral amplitude $\exp(-l(x,y))$ used for the definition of  the 
Euclidean propagator  is modified so that it is invariant under the duality 
transformation $l\to L_P^2/l$, then (i) the propagator becomes UV finite and 
(ii) the result is identical to that obtained by changing the interval 
$(x-y)^2$ to $(x-y)^2+4L_P^2$. This shows a deep connection between the 
duality  $l\to L_P^2/l$ and the existence of zero-point length, which was not 
clearly understood. 

The string theory has both these features (T-duality as well as finite length 
scale) built into it in a natural fashion. This suggests that one should be 
able to obtain similar results in standard 
string theory.
In this letter we  show that this is indeed the case: the 
 zero-point length can be reproduced using $T$-duality of the string theory and
 the leading order corrections to 
the propagator, defined in a specific but consistent way, is the same as that 
obtained from the principle of path
integral duality used in  \cite{padma2}.

Since it is difficult 
(~and unnecessary for our purpose~) to obtain the low energy
quantum field theory description directly from {\it full} string theory, we 
shall merely invoke certain key ingredients of the string theory to achieve 
our goal. There are three technical issues which are relevant to  establishing
 such a  correspondence
between the language of string theory and the low energy quantum field theory, 
so that one can   
sensibly ask for the corrections to the low energy propagator from string 
theory. The first point is that
the result we want to derive is non-perturbative so that the simplest
procedure, of doing this in a perturbative manner, is inadequate. The 
corrections to the propagator from zero-point length typically involves terms 
like $[x^2+L_P^2]^{-1}$.
This quantity is finite as $x\to 0$ but if we expand it in powers of 
$L_P^2/x^2$, each term of the expansion will diverge as $x\to 0$. So we need 
to approach the problem non-perturbatively. Second,  we need to identify
a suitable (collective) degree of freedom of the string theory, which could be 
thought of as analogous to the low energy concept of a particle which is 
propagating. We find that the string center of mass (SCM) is a suitable 
candidate for this role and one can define a path integral propagator for the 
same. This propagator is UV finite because of the winding modes in the compact 
dimensions.  Finally, we also need a mechanism to incorporate the light
degrees of freedom (with masses far lower than the Planck mass) so that one 
can talk about a propagator for, say,
an electron. This is possible by incorporating the effects of light modes by 
adding a suitable mass term in the Schwinger's proper time formulation of the 
propagator. We shall now present the details of this approach.

Closed strings have the unique property to wrap around compactified 
 spatial dimensions. In the case of a single compact dimension, the string 
 mass spectrum is labeled, in addition to the harmonic excitation occupational 
 number, by two more quantum numbers. The number $n$ which determines the
  Kaluza-Klein excitation level and the  number $w$ which counts 
  the string winding around compactified dimensions. The string mass spectrum  
  is given by \cite{book}:
  \begin{equation}
    M^2 =\frac{1}{2\alpha^\prime}\left(\, n^2 \frac{\alpha^\prime}{R^2} + 
    w^2\frac{R^2}{\alpha^\prime}\, \right)+ \dots \label{M}
     \end{equation}
  where, $1/\alpha^\prime$ and $R$ are the string tension and the
 compactification radius respectively. (The dots refer to harmonic
 oscillators contributions which are irrelevant for our path integral 
 calculation;
some relevant features of the low energy excitations will be incorporated at 
the end of the calculation indirectly.)
 For the sake of simplicity, we have written (\ref{M}) for the case of a
 single compact dimension (~with the fifth-dimension compactified
 on a circle of length $l_0=2\pi R$~) but our analysis can be extended
 to the case of $D$ dimensional hyper-torus in a straightforward way. (We
quote the result for the general case in the end).
 The important property of the spectrum (\ref{M}) is its invariance
 under so-called $T$-duality. It is defined as a transformation  relating 
 the compactification radius and its inverse, while, at the same time,
 exchanging quantum numbers $n$ and $w$ as:
 $R \longleftrightarrow \alpha^\prime /R$, $n \longleftrightarrow w$.
This  symmetry  relates large and small
 distances, and  there is also a unique length  $\sqrt{\alpha^\prime}$ that is 
 invariant under $T$-duality which is a possible candidate for the 
 \textit{zero-point length}.
 
Let us consider a \textit{closed} string winding $w$ times around the
compact dimension and  freeze all the oscillator modes.
 By dropping the oscillator modes,
  we have reduced string dynamics to the motion of its center of mass.
 We will quantize the dynamics of the string center of mass (SCM), 
 i.e. the Kaluza-Klein and winding modes. This is appropriate since we 
 eventually want to obtain a
corrected Feynman propagator in field theory. The latter can be intuitively 
thought of as giving the amplitude
of propagation for a ``particle'', the closest analogue of which is the string 
center of mass.
We start from  the  propagation kernel of SCM in $4+1$ dimensions  defined as
 
\begin{eqnarray}
&&K\left[\, x-y\ ; T\,\right]=\nonumber\\
&&\sum_{n=-\infty}^\infty
\int_{z(0)=x}^{z(T)=y}
\int_{x^5(0)=0}^{x^5(T)=n\,l_0}\left[\,Dz\,\right] \left[\,Dp\,\right]
\left[\, Dx^5\,\right]
 \left[\,Dp_5\,\right]\times\nonumber\\
&&\exp\left[\, i\int_0^T d\tau\left(\,p_\mu\dot x^\mu +p_5\dot x^5
-\frac{i}{2\mu_0}\left(\, p_\mu p^\mu +  p_5 p^5\,\right)\,\right)\,\right]
\nonumber\\
&& \label{kernel}
\end{eqnarray}   
where, $P_M\equiv \left(\, p_\mu\ , p_5\,\right)$ is the SCM five-momentum
and $\mu_0$ is  a  parameter introduced for dimensional reason which
will not appear in the final result. In (\ref{kernel})
we are summing over histories which in four dimensional spacetime
connect $x$ and $y$ in  time $T$ while, at the same time, they wind 
$n$ times around $x^5$.

In order to obtain the four dimensional propagator we need to
integrate out the fifth-component of the paths. This is most easily done
 by writing
\begin{equation}
\int_0^T p_5 \dot x^5 d\tau 
= \left[\,p_5\,x^5\,\right]_0^T
-\int_0^T d\tau x^5\dot p_5\label{p5}
\end{equation}
and computing the path integral by
\begin{equation} 
\int\left[{\cal D}x^5\right]\exp\left(-i\int _0^T d\tau\,
x^5\,\dot p_5 \,\right)=\delta\left[\,\dot p_5\,\right] 
\label{x5}
\end{equation}
This shows that $p_5$ is $\tau$-independent, as it should be for a ``free''
particle. The final effect is to reduce functional
integration over $p_5$ to an ordinary integral \cite{ej}, but the
 compactification of the extra-dimension imposes a quantization
condition on the SCM $p_5$ component of momentum, as: $p_5=n \, \mu_0 (l_0/T)$ 
and we need to sum over the winding number $n$. This constraint can be 
implemented by introducing a Dirac-delta functional $\delta[p_5-n \mu_0
(l_0/T)]$ in the path integral. Doing this and evaluating the integrals over 
the $p_5$ we get:   
\begin{eqnarray}
&&K\left(\, x-y\ ; T\,\right)=2\sum_{n=1}^\infty
\exp\left(\frac{in^2\mu_0 l_0^2}{2T}\right)
\times\nonumber\\
&&\int_{z(0)=x}^{z(T)=y} \left[{\cal D}z\right] \,\left[{\cal D}p\right]\,
\exp\left(\, i\int_x^y p_\mu\, dz^\mu -i\int_0^T d\tau \frac{p^\mu\, p_\mu}
{2\mu_0}\,\right)\nonumber\\
&&
\end{eqnarray}
 In the  summation over $n$, the $n=0$ mode can be dropped since
it does not sense the effect of the compact dimension which we are interested 
in. Further since
only $n^2$ appears, the sum over negative integers is the same as that over 
positive integers.
These features are already incorporated in the above result.
The memory of the compact dimension is left over in the exponent involving
summation over Kaluza-Klein modes.

The remaining integration over the
four dimensional components can be done in a similar manner. Using

\begin{eqnarray}
\int_{z(0)=x}^{z(T)=y}\left[{\cal D}z\right]&& \exp\left(\, i\int_x^y d\left(\,
p_\mu z^\mu\,\right)-i\int_0^T d\tau\, z^\mu \dot  p_\mu\,\right)=\nonumber\\
 && \delta\left[\,\dot p_\mu\,\right]
\exp i\left[\,p_\mu z^\mu \,\right]_x^y
\end{eqnarray}
and integrating over $p_\mu$, we get the final result:
\begin{eqnarray}
K_{reg}&&\left(\, x-y\ ; T\,\right)=
-2\left(\, \frac{\mu_0}{2i\pi\, T}\,\right)^{2}\times
\nonumber\\
&&  \sum_{n=1}^\infty\exp \left\{\, i\frac{\mu_0}{2T}
\left[\, \left(\,x-y\,\right)^2 +n^2l_0^2\,\right]\,\right\}
\label{Kreg}
\end{eqnarray}
The global normalization factor of the path integral measure is set in such a 
way  to produce the 
correct form of the propagator  in  the limit $l_0\to 0$.
 
The leading correction comes from the $n=1$ which winds around the compact 
dimension just once. This reduces the expression in 
 (\ref{Kreg}) to exactly the result obtained in \cite{padma2} from path integral
 duality. Thus the hypothesis introduced in  \cite{padma2}, viz. the 
 modification
\begin{equation}
\sum\exp[-l(x,y)]\to \sum\exp\left[-l+\frac{l_0^2}{l}\right]
\end{equation}
captures the leading order corrections from string theory.
The origin of zero-point length can be traced to  the existence of compact 
extra-dimension. The ultraviolet divergences are, of course, cured because of 
the length scale $l_0$.

Let us next compute $G\left(\, x-y\,\right)$ from
the kernel by integrating over all possible values of the time lapse $T$ with 
proper weight.
In the standard Schwinger proper time approach to the propagator, this is done 
by using the weight
$\exp (-im^2T)$ where $m$ is the physical mass of the particle which will 
emerge from the pole of the propagator.
 In our case, the SCM has a mass equal to the string tension times the total 
length of the string wrapped $w$ times along $x^5$  , i.e.
$M_{SCM}=w \, l_0/\alpha^\prime $ (~In order to avoid
any confusion, we remark that $n$ counts the winding of the SCM histories
summed in the path integral, while $w$ refers to the  string wrapping.~) Thus, 
the Green function of SCM will be as
\begin{eqnarray}
G_{reg}\left(\,x-y\,\right)&&\equiv \frac{i}{ 2\mu_0}\sum_w 
\int_0^\infty dT \exp\left(\frac{iT}{2\mu_0}\, m_0^2\,\right)\times\nonumber\\
&&\exp\left(\frac{iT}{2\mu_0}\frac{w^2 l_0^2}{\alpha^{\prime\, 2}}\,\right)
 K_{reg}\left(\, x-y\ ; T\,\right) \label{g}
\end{eqnarray}
where $m_0$ is the mass of the particle in the limit $l_0\to 0$, i.e. is the
mass measured at low energies compared to  threshold energy of the KK modes. 
The summation over
$w$ takes care of the KK modes and we have introduced $m_0$  to facilitate 
comparison with results of low energy field theory. (It can be set to zero for 
obtaining purely the SCM propagator).  
Inserting (\ref{Kreg}) into  (\ref{g}) we find

\begin{eqnarray}
&& G_{reg}\left(\,x-y\,\right)=\frac{1}{(2\pi)^2}\sum_{ w , n=1}^\infty
\int_0^\infty  \frac{ du}{u^2}\exp\left(\frac{i\,u}{2}\, m_0^2\,\right)
\times\nonumber\\
&&\exp\left( \frac{iu}{2}\frac{w^2 l_0^2}{\alpha^{\prime\, 2}} \right)
 \exp \left\{\,\frac{i}{2u}
\left[\, \left(\,x-y\,\right)^2 +n^2l_0^2\,\right] \,\right\}
\end{eqnarray}
the unobservable parameter $\mu_0$ has been eliminated by suitable
rescaling $T\longrightarrow u\equiv T/\mu_0$.
The integral can be expressed in terms of the Bessel functions:

\begin{eqnarray}
&& G_{reg}\left(\,x-y\,\right)=\nonumber\\
&& \frac{1}{(2\pi)^2}\sum_{ w , n=1}^\infty
\sqrt{\,\frac{m_0^2 + w^2 l_0^2/\alpha^{\prime\, 2}}
{\left(x-y\,\right)^2 +n^2\,l_0^2}}
\times\nonumber\\
&&K_1\left(\, 
\sqrt{\,\left[\, \left(x-y\,\right)^2+n^2\,l_0^2\,\right]\left[\, 
  m_0^2 + w^2 l_0^2/\alpha^{\prime\, 2} \,\right]}
\,\right)
\label{final}
\end{eqnarray}
where $K_1(z)$ is a modified Bessel function. Inspection of (\ref{final}) 
shows that $G_{reg}\left(\,x-y\,\right)$ is invariant under the transformation
$(u/\alpha^\prime)\longleftrightarrow (\alpha^\prime/u)$, $n 
\longleftrightarrow w$.
This is the same as the T-duality (~which makes  the spectrum (\ref{M}) invariant
under  $R \longleftrightarrow \alpha^\prime /R$, $n \longleftrightarrow w$~)
with the identifications 
\begin{equation}
(\alpha^\prime/R^2)=(l_0^2/u);\quad 
(R^2/\alpha^\prime)=(u l_0^2/\alpha^{\prime\, 2})
\end{equation}
 which, in turn, requires 
precisely the identification $l_0^2=\alpha^\prime$. That is, the  minimum 
length scale in string theory, set by  $\sqrt{\alpha^\prime}$, does get 
identified with the zero-point length, $l_0 \equiv \sqrt{\alpha^\prime}$,
when $T$-duality is properly encoded into the two-point Green function 
$G\left(\, x-y\,\right)$. This explains the
$T$-duality origin of the zero-point length, in four dimensional spacetime.

The leading term in (\ref{final}) is the one corresponding to $n=w=1$. 
Writing $m^2\equiv m_0^2 + l_0^2/\alpha^{\prime\, 2}$, we get the leading 
terms to be
\begin{eqnarray}
&& G_{reg}\left(\,x-y\,\right)\approx
 \frac{1}{(2\pi)^2}\left[\, 
\frac{m}{\,\sqrt{\,\left(x-y\,\right)^2 +n^2\,l_0^2}}\right. \times\nonumber\\
&& \left. K_1\left(m\sqrt{\,\left(x-y\,\right)^2
+n^2\,l_0^2}\,\right)\,\right]
\end{eqnarray}
The Fourier transform of this term gives the propagator in the momentum space,
which is
\begin{equation}
 G_{reg}\left(\,p^2\,\right)\approx
 \frac{1}{(2\pi)^2} 
\frac{l_0 K_1\left(\, l_0\sqrt{\,p^2+m^2}\,\right) }{\,\sqrt{\,p^2 + m^2}}
\end{equation}
which is the result found earlier in  \cite{padma2}.

Strictly speaking, the bosonic string can be consistently described only
$26$ dimensions.  This requires the extension of $4+1$ calculations to
$4+22$ dimensions. In a toroidal compactification scheme
generalization of our results is easily achieved. There are compact dimensions
of a priori different  radii $R_i$ and corresponding Kaluza-Klein and winding
modes $n_i$ and $w_i$. Nevertheless $T$-duality will relate all of them
to the single parameter in the theory which is the string length
$\sqrt{\alpha^\prime}$. Thus,  $T$-duality guarantees a unique zero-point
length in the theory. Therefore, Green function can be 
obtained from (\ref{final}) by the introduction of additional sum over
compactified dimensions as:
\begin{equation}
 n^2\, \longrightarrow \sum_{i=1}^{22} n_i^2\ ;\quad
w^2  \longrightarrow \sum_{i=1}^{22} w^2_i
\end{equation}

We summarize our results as follows: Starting from the low energy end, 
we have the standard formulation of quantum field theory in which one can 
associate a propagator with a particle. Naively speaking, this propagator, 
$G(x-y)$ gives the amplitude for the particle to propagate from $x$ to $y$. 
The ultra violet divergences of the standard quantum field theory can be 
related to the fact that, $G$ diverges as $x\to y$. One expects quantum 
gravitational effects to cure this behaviors and make $G$ ultraviolet finite 
if there is a fundamental length scale in the theory. It was shown in 
\cite{padma2} that such a result can be obtained if one postulates that the 
path integral amplitude used to compute $G$ is invariant under the duality 
prescription $l\to L_P^2/l$. However, it was not clear how the duality and 
zero-point length of the spacetime (which leads to UV finiteness) are related. 
In this work, we have approached the problem from the string theory end, using 
the fact that
standard formulation of string theory has
both a fundamental length scale and T-duality. It is, however, not easy to 
obtain the low energy predictions
of the string theory or the effective propagators for the low energy 
excitations, directly from the currently available formulation. This is because 
the  low energy description of 
different particles is expected to emerge through a complex process in full 
fledged string theory. Though this makes an exact calculation difficult,  one 
can  capture 
 the essential physics by treating the string center of mass as the analogue of 
 the particle in the low energy limit and compute a propagator for it. We found 
 that the winding over the compact dimensions leads to an infinite sum in this 
 propagator. The $n=w=1$ term gives the leading order contribution, which is 
 identical to the result obtained in \cite{padma2} using the prescription of 
 path integral duality. This shows that, to the lowest order, the effects of 
 string theory can be incorporated into ordinary quantum field theory by 
 replacing $(x-y)^2$ by $(x-y)^2+4L_P^2$ in the Schwinger
proper-time Kernel. 

We incorporated the effect of low energy excitations 
by adding the corresponding phase term with a mass to the Schwinger proper-time 
integral. This allows us to define
a low energy propagator and compare it with the results of earlier work. We 
stress that the incorporation of the low mass excitations is logically 
separate from the calculation of propagator for SCM and the results for pure 
SCM propagator can be obtained by setting $m_0=0$.

The corrections to the standard path integral propagator obtained here 
[as well as in the earlier work] are non-perturbative by nature. That is, they 
cannot be obtained by a Taylor series expansion in $L_P^2$. The expressions 
involved are like $[x^2+L_P^2]^{-1}$ which is finite as $x^2\to 0$; but if it 
is expanded in powers of $L_P^2/x^2$, each term in the expansion will diverge 
as $x\to 0$.

One of the authors (T.P) thanks the Institute of Astronomy, Cambridge for 
hospitality where part of the work was carried out.

\end{document}